\documentclass[twocolumn,prd,aps,showpacs,tightenlines,nofootinbib,preprintnumbers, 
floats]{revtex4}
\usepackage{epsfig}
\usepackage{graphicx}
\usepackage{dcolumn}
\usepackage{bm}
\usepackage{amssymb}
\usepackage{amsmath}
\usepackage{amsbsy}

\newcommand{\bomega}{\mbox{\boldmath$\omega$}}
\newcommand{\RRdual}{\mbox{\boldmath$ R \tilde R$}}
\newcommand{\Rdual}{\mbox{$\tilde R$}}

                              \newlength{\strikewidth}
                              \newlength{\strikelength}
\begin{document}

\title{The effects of Chern-Simons gravity on bodies orbiting the earth}

\author{Tristan L.~Smith$^1$}\email{tlsmith@tapir.caltech.edu}
\author{Adrienne L.~Erickcek$^1$} \email{erickcek@tapir.caltech.edu}
\author{Robert R.~Caldwell$^2$}
\author{Marc Kamionkowski$^1$}
\affiliation{$^1$California Institute of Technology, Mail Code
130-33, Pasadena, CA 91125}
\affiliation{$^2$Department of Physics \& Astronomy, Dartmouth
College, 6127 Wilder Laboratory, Hanover, NH 03755}


\begin{abstract}
One of the possible low-energy consequences of string theory is the addition of a Chern-Simons term to the standard Einstein-Hilbert action of general relativity.  It can be argued that the quintessence field should couple to this Chern-Simons term, and if so, it drives in the linearized theory a parity-violating interaction between the gravito-electric and gravitomagnetic fields.  In this paper, the linearized spacetime for Chern-Simons gravity around a massive spinning body is found to include new modifications to the gravitomagnetic field that have not appeared in previous work. The orbits of test bodies and the precession of gyroscopes in this spacetime are calculated, leading to new constraints on the Chern-Simons parameter space due to current satellite experiments.
\end{abstract}

\maketitle

\section{Introduction}

The study of modifications of the theory of general relativity has been of interest ever since Einstein first formulated general relativity in 1915.  Particularly interesting are modifications that introduce terms to the Einstein-Hilbert action that are second order in the curvature, as such modifications represent high-energy corrections to the Einstein-Hilbert action that might arise in quantum gravity. Chern-Simons gravity is an example of such a second-order modification of the Einstein-Hilbert action.  

Chern-Simons modifications to gravity were first considered in 2+1 dimensions \cite{deser}.  Refs.~\cite{campbell,campbell2} investigated the structure of these theories in 3+1 dimensions and showed how they could arise as a low-energy consequence of string theory.  Ref.~\cite{lue} considered some early-universe implications of such theories.  Refs.~\cite{alexander_peskin} investigated how Chern-Simons terms might participate in leptogenesis.  Ref.~\cite{jackiw} renewed the investigation of Chern-Simons gravity, working out the linearized equations of the theory and their implications for gravitational waves. Most recently, Refs.~\cite{alexander} solved the linearized Chern-Simons field equations around a collection of spinning point masses.  In much of the work on Chern-Simons gravity, the Chern-Simons term is coupled to a scalar field (as detailed below), and this scalar field is assumed to be spatially homogeneous but time varying.  This assumption can be motivated by arguments analogous to those that have been made suggesting that the quintessence field should be coupled to the Chern-Simons term of electromagnetism \cite{sean}.

Chern-Simons gravity has thus far eluded constraints from Solar System tests of weak-field gravity because it is indistinguishable from general relativity for all spacetimes  that possess a maximally symmetric two-dimensional subspace and for all conformally flat spacetimes \cite{campbell}.  Therefore, the Schwarzschild spacetime as well as the Robertson-Walker spacetime are also solutions of the Chern-Simons gravitational field equations.  Distinguishing Chern-Simons gravity from general relativity requires considerations of spacetimes that are not spherically symmetric, such as the spacetime around a spinning body.  To this end, Refs.~\cite{alexander} investigated the Chern-Simons modifications to the motion of bodies around a spinning point mass and found that the motion was indistinguishable from that in general relativity. 

In this paper we take further steps to link Chern-Simons gravity to current and forthcoming experimental tests of weak-field gravity.  We assume, as in other recent work, that the scalar field coupled to the Chern-Simons term is time varying but spatially homogeneous.  We then determine the spacetime around an extended spinning mass and find that it differs from the spacetime around a spinning point mass.  We determine the orbits of test particles and the precession of gyroscopes moving in this spacetime and find that the Chern-Simons modification does lead to observable deviations from the predictions of general relativity.  These deviations allow us to evaluate constraints to the Chern-Simons parameter space from current satellite experiments, as well as those regions to be probed with forthcoming experiments.

The paper is organized as follows:  Section II defines the theory and derives the gravitational field equations.
Section III considers the linear theory and derives the gravitomagnetic equations of motion (the Chern-Simons Amp\`ere's law).   Section IV discusses the solution for the gravitomagnetic field around a spinning massive body.  In Section V, we consider the orbital precession of test bodies in this spacetime, as well as the orbital precession of gyroscopes, and we determine the regions of the Chern-Simons-gravity parameter space that are probed with the LAGEOS and Gravity Probe B satellites.  We conclude briefly in Section VI.  Appendix A shows how the Chern-Simons Lagrangian we work with may be derived from a string-theory action, and Appendix B outlines the derivation of the gravitomagnetic vector potential around a spinning sphere.
\section{CHERN-SIMONS GRAVITY}

We consider the theory defined by the action
\begin{eqnarray}
     S &=& \int {\rm d}^4 x \sqrt{-g} \left[-\frac{1}{2\kappa^2} R+
     \frac{\ell}{12} \theta \RRdual - \frac{1}{2} (\partial
     \theta)^2 \right. \\ \nonumber
     & & \left. - V(\theta) +{\cal L}_{\mathrm{mat}}\right],
\end{eqnarray}
where ${\cal L}_{\mathrm{mat}}$ is the Lagrangian density for matter, $g\equiv \det g_{\mu\nu}$ is the determinant of the metric $g_{\mu\nu}$, $R$ is the Ricci scalar (with the convention $R^{\lambda}_{\ \ \mu \nu \kappa} \equiv \Gamma^{\lambda}_{\mu \nu, \kappa} + \cdots$ for the Riemann tensor), and $\RRdual$ is a contraction of the Riemann tensor and its dual:
\begin{equation}
     \RRdual \equiv R^{\beta\ \gamma \delta}_{\ \alpha}
     \tilde{R}^{\alpha}_{\ \beta \gamma
\delta}, 
\end{equation}
where the dual of the Riemann tensor is defined by
\begin{equation}
     \Rdual^{\mu}_{\ \nu \alpha \beta} \equiv \frac{1}{2}
     \epsilon_{ \sigma \tau \alpha \beta}  R^{\mu \ \ \sigma \tau}_{\ \ \nu},
\end{equation}
where $\epsilon_{ \sigma \tau \alpha \beta}$ is the Levi-Civita tensor, including a factor of $\sqrt{-g}$.  Finally, $\ell$ is a new length scale, a parameter of the theory, and $\kappa^2 \equiv 8 \pi G$, where $G$ is Newton's constant.  Throughout this paper we take Greek indices to range from 0 to 3.  Appendix A shows how such an action may arise in string theory.  This action is different from the action considered in Ref.~\cite{jackiw} in that here $\theta$ is a dynamical scalar field with a canonical kinetic term, so the $\ell$ parameter is required to make the action dimensionless.

The equation of motion for $\theta$ is given by
\begin{equation}
     \Box \theta = \frac{{\rm d} V}{{\rm d}\theta}-\frac{1}{12} \ell \RRdual .
\label{scalar}
\end{equation}
The gravitational field equations take the form 
\begin{equation}
     G_{\mu \nu} - \frac{2}{3} \ell \kappa^2 C_{\mu \nu} =
     -\kappa^2 T_{\mu \nu}, 
\label{field_equation}
\end{equation}
where $G_{\mu\nu}$ is the Einstein tensor, $T_{\mu\nu}$ is the
stress-energy tensor for the scalar field and the matter Lagrangian, and we
refer to $C_{\mu \nu}$ as the Cotton-York tensor\footnotemark,
\begin{eqnarray}
     C^{\mu \nu} &=& \frac{1}{2} \bigg[ (\partial_{\sigma} \theta) \left(
     \epsilon^{\sigma \mu \alpha  
     \beta} \nabla_{\alpha} R^{\nu}_{\beta} 
     + \epsilon^{\sigma \nu \alpha \beta} \nabla_{\alpha}
     R^{\mu}_{\beta}\right) \label{cy}\\ \nonumber
&+& \nabla_{\tau} (\partial_{\sigma} \theta) \left( \tilde R^{\tau \mu \sigma
     \nu} + \tilde R^{\tau \nu 
\sigma \mu}\right)\bigg].
\end{eqnarray}
 Appendix A
provides an alternative expression for the Cotton-York tensor.

Ref.~\cite{jackiw} notes that if $\theta$ is a non-dynamical field (a
Lagrange multiplier), the theory cannot accommodate a spacetime
with a nonzero $\RRdual$ because the Cotton-York tensor would have a non-zero divergence.
 However, if $\theta$ is a dynamical field, then the theory
can indeed accommodate spacetimes with nonzero $\RRdual$ since we have
\begin{equation}
     -\frac{2}{3} \ell \kappa^2 \nabla^{\mu} C_{\mu
     \nu} = \frac{\ell \kappa^2}{12} (\partial_{\nu} \theta) \RRdual =  -\kappa^2  \nabla^{\mu}T^{\theta}_{\mu \nu},
\label{conserv_se}
\end{equation}
where $T^{\theta}_{\mu \nu}$ is the stress-energy tensor for $\theta$. 
We see that whereas the scalar-field stress-energy and the Cotton-York tensors are
separately conserved when $\RRdual=0$, 
the divergence of the scalar field stress-energy tensor is
precisely balanced by the
divergence of the Cotton-York tensor for non-zero $\RRdual$ due to the novel
coupling between the scalar field and gravity. 
\section{The Chern-Simons gravitomagnetic equations}

We begin with a perturbation to the flat metric [using signature
$(-+++)$], 
\begin{equation}
     g_{\mu \nu} = \eta_{\mu \nu} + h_{\mu \nu}, 
\end{equation}
and compute the linearized Einstein and Cotton-York tensors,
\begin{widetext}
\begin{eqnarray}
     G_{\mu \nu}^{\mathrm{linear}} &=& \frac{1}{2}(\Box h_{\mu
     \nu} + \partial_{\mu}\partial_{\nu} h -
     \partial_{\mu}\partial_{\alpha} h^{\alpha}_{\nu} - 
     \partial_{\nu} \partial_{\alpha} h^{\alpha}_{\mu} - 
     \eta_{\mu \nu}[\Box h - \partial_{\alpha} \partial_{\beta}
     h^{\alpha \beta}]), \\ 
     C_{\mu \nu}^{\mathrm{linear}} &=& \frac{1}{8}
     \partial^\alpha \partial_\beta 
     \theta [\eta_{\nu \gamma} \epsilon^{\gamma\beta\sigma\tau}\left({h_{\mu 
     \sigma},_{\alpha\tau}-h_{\alpha\sigma},_{\mu\tau}-
     h_{\mu\tau},_{\alpha\sigma}+h_{\alpha\tau},_{\mu
     \sigma}}\right) +\eta_{\mu 
     \gamma} \epsilon^{\gamma\beta\sigma\tau}\left({h_{\nu
     \sigma},_{\alpha\tau}-
     h_{\alpha\sigma},_{\nu\tau} - h_{\nu\tau},_{\alpha\sigma} +
     h_{\alpha\tau},_{\nu \sigma}}\right)]\nonumber \\
      &+& \frac{1}{4}\partial_\beta \theta 
      \epsilon^{\alpha\beta\sigma\tau}
     [\eta_{\alpha\mu}\partial_\tau \left({\Box
     h_{\nu\sigma}-\partial_\nu \partial^\lambda
     h_{\lambda\sigma}}\right)
     +\eta_{\alpha\nu}\partial_\tau\left({\Box
     h_{\mu\sigma}-\partial_\mu \partial^\lambda 
     h_{\lambda\sigma}}\right)],
\end{eqnarray}
\end{widetext}
 \footnotetext{We note that this definition differs from the usual expression\\ for the four-dimensional Cotton-York tensor (see Ref.~\cite{jackiw}).}
where $\Box$
 is the flat-space 
d'Alembertian and the comma denotes partial differentiation.  
Since we will require below only the gravitomagnetic fields, we will be primarily interested in the time-space
components of 
the linearized field equations. 

In this paper, we suppose that the scalar field depends only on cosmic time, $\theta=\theta(t)$, the assumption being that $\theta$ is either a quintessence field or some other field that somehow echoes the arrow of time associated with the cosmic expansion.  This choice implies that the field equations are not Lorentz invariant in the Solar System since $\partial_{\sigma} \theta$ points in the cosmic time direction and couples to local gravity through the Cotton-York tensor [Eq.~(6)].   We note that a nonzero $\RRdual$ will source spatial variations in $\theta$ through Eq.~(4).  By restricting $\theta$ to be spatially homogenous, we are effectively treating $\theta$ as a nondynamical field, and we leave a full dynamical treatment to future work.  Finally, we neglect corrections due to the motion of the Earth with respect to the rest frame of the cosmic microwave background.

We work with the trace-reversed metric perturbation,
 \begin{equation}
     \bar{h}_{\mu \nu} \equiv h_{\mu \nu} -\frac{1}{2} \eta_{\mu \nu} h,
\end{equation}
and impose the Lorenz-gauge condition, $\partial^{\mu} \bar{h}_{\mu \nu} = 0$, to obtain the linearized time-space
field equations,
\begin{eqnarray}
     G_{0 i}^{\mathrm{linear}} - \frac{2}{3} \ell \kappa^2C_{0
     i}^{\mathrm{linear}} &=& -\kappa^2 T_{0 i},
\end{eqnarray}
with
\begin{eqnarray}
     G_{0 i}^{\mathrm{linear}} &=& \frac{1}{2}\Box \bar{h}_{0
     i}, \\
     C_{0 i}^{\mathrm{linear}} &=& \frac{
     \dot{\theta}}{4}\epsilon^0_{\ i j k} 
     \partial^{j}\Box \bar{h}_{\ 0}^{k},
\end{eqnarray}
where the dot denotes differentiation with respect to time and Latin indices are purely spatial and range from 1 to 3. 
The stress-energy tensor for $\theta(t)$ is diagonal, so it does not contribute to the time-space field equations.

Let $t^{\alpha}$ be a unit vector in the coordinate time direction, and then define the 4-vector potential of this
linearized theory,
\begin{equation}
     A_{\mu} \equiv - \frac{1}{4} \bar{h}_{\mu \nu} t^{\nu} = -
     \frac{1}{4} \bar{h}_{\mu 0} .
\label{vec_pot_def}
\end{equation}
We consider a source with mass density $\rho$, mass current $\vec{J}$ and negligible pressure, so we can express the matter stress-energy tensor as 
\begin{equation}
     T_{\mu \nu} = 2 t_{(\mu} J_{\nu)} - \rho t_{\mu} t_{\nu},
\end{equation}
where $J_\mu \equiv -T_{\mu\nu}t^\nu = (-\rho, \vec{J})$.  In general relativity, the time-space components of the linearized field equations take the form
\begin{equation}
     \partial^{\mu} \partial_{\mu} A_{i}= -4 \pi G J_{i},
\end{equation}
which is (nearly) identical to Maxwell's equations for the vector potential in Lorenz gauge ($\partial_\mu A^\mu = 0$). Given our definition of $A^\mu$, the Lorenz-gauge condition for $A_\mu$ is implied by our earlier gauge choice for
$\bar h_{\mu\nu}$.  

The classically `physical' fields (i.e., those that enter into the geodesic equation) $\vec{E}$
and $\vec{B}$ are given by
\begin{eqnarray}
     E^i &=& \partial_i A_0 - \partial_0 A_i, \\
     B^i &=& \epsilon^{0ijk}\partial_j A_k,
\end{eqnarray}
where we have defined $\epsilon^{0ijk} = 1$.
Two of the Maxwell equations,
\begin{eqnarray}
     \vec{\nabla} \cdot \vec{B} &=& 0,\\
     \vec{\nabla} \times \vec{E} &=& - \frac{\partial
     \vec{B}}{\partial t},
\end{eqnarray}
are a direct consequence of the way in which the $\vec{E}$ and $\vec{B}$ fields are defined in terms of the vector potential, and so these two equations will be the same in Chern-Simons gravity.  Gauss' law, which follows from the time-time component of the field equation, is now
\begin{equation}
     \vec{\nabla} \cdot \vec{E} = 4 \pi G (\rho + \rho_\theta)
\end{equation}
where $\rho_\theta$ is the energy density of the scalar field $\theta(t)$ and is uniform throughout the Solar System.  Since $\rho_\theta$ cannot be larger than the mean cosmological energy density, it must be negligible compared to the density of the source $\rho$, and we do not consider it further.   The only significant modification will be to Amp\`ere's law, which, for Chern-Simons gravity, is now given by
\begin{equation}
     \vec{\nabla} \times \vec{B} - 
     \frac{\partial \vec{E}}{\partial t} - \frac{1}{m_{\rm cs}}\Box \vec{B} = 4 \pi G \vec{J},
\label{amp}
\end{equation}
where we have defined $m_{\rm cs} \equiv -3/(\ell \kappa^2 \dot{\theta})$.

Given the metric perturbation represented by the gravitomagnetic potential and neglecting the time variation of the metric, slowly moving particles travel on geodesics such that a `Lorentz force law' of the form,
\begin{equation}
   \vec{a} = - \vec{E} - 4 \vec{v} \times \vec{B},
\label{force}
\end{equation}
is obtained.  Therefore, as in electrodynamics, only the physical fields, and not the potentials, have physical relevance.

We furthermore note that $\RRdual$ can be expressed in terms of gravito-electric
and gravitomagnetic fields as
\begin{equation}
\RRdual = -16 (\partial_i E_j) (\partial_k B_l) ( \eta^{ik}\eta^{jl} + \eta^{il}\eta^{jk}).
\end{equation}
Unlike the case with Maxwell fields \cite{carroll}, it is not
sufficient for the fields to have a non-vanishing $\vec{E} \cdot
\vec{B}$ in order to have a non-trivial coupling between gravity
and the scalar field. The best example of a gravitational source
which produces a non-vanishing $\RRdual$ is a spinning,
spherical body.

\section{Gravitomagnetism due to a Spinning Sphere in Chern-Simons gravity}
\label{sec:Bfield}

We are now in a position to calculate the gravitomagnetic field in Chern-Simons gravity for a spinning body.  Appendix B provides details of the calculation. 

We consider a homogeneous rotating sphere, and so the mass current is
\begin{equation}
     \vec{J} = \rho \left[\vec{\omega} \times \vec{r}\right] \Theta(R-r),
\end{equation}
where $R$ is the radius of the rotating body, $\rho$ is its density, $\vec{\omega}$ is its angular velocity,  $r$ is the distance from the origin, and $\Theta$ is the Heaviside step function.  As detailed in Appendix B, the field equation, Eq.~(\ref{amp}), is rewritten as an equation for $\vec{A}$ and is solved by imposing the condition that the metric be continuous everywhere and that the gravitomagnetic field be finite and well-behaved at the origin; the resulting vector potential is given in Appendix B.  We note that in deriving this solution we have assumed that the time derivative of $m_{\rm cs}$ is negligible.  The gravitomagnetic field is then obtained by taking the curl of $\vec{A}$ and may be written as $\vec{B} =  \vec{B}_{\rm GR}+ \vec{B}_{\rm CS}$, where
\begin{eqnarray}
     \vec B_{\rm GR}  = &&\frac{4 \pi G \rho R^2}{15}\\ 
     &&\times 
\begin{cases}
     \left(5-3\frac{r^2}{R^2}\right) \vec\omega + 3\frac{r^2}{R^2} \hat r \times (\hat r \times \vec
     \omega), \ &r
        \leq R, \\
     \frac{R^3}{r^3}
     \left[2 \vec\omega + 3 \hat r \times (\hat r \times \vec
     \omega)\right],\ &r \geq R, \nonumber
\end{cases}
\label{BfullGR}
\end{eqnarray}
is the gravitomagnetic field inside and outside a spinning sphere in general relativity, and
\begin{eqnarray}
     \vec B_{\rm CS} &=& 4 \pi G \rho R^2\left\{ D_1(r)\, \vec\omega
     +  D_2(r)\, \hat r \times \vec\omega \right. \nonumber \\
     &&\left. + D_3(r) \, \hat r \times (\hat r \times
     \vec\omega) \right\},
\label{BfullCS}
\end{eqnarray}
is the new contribution in Chern-Simons gravity.  Inside the sphere $(r\leq R)$, 
\begin{eqnarray}
    D_1(r) &=& \frac{2}{(m_{\rm cs}R)^2}+\frac{2R}{r}\, y_2 (m_{\rm cs}R) j_1 (m_{\rm cs}r),  \nonumber\\          
    D_2(r) &=& \frac{m_{\rm cs} r}{(m_{\rm cs}R)^2}+m_{\rm cs}R\, y_2 (m_{\rm cs}R) j_1 (m_{\rm cs}r), \nonumber\\
    D_3(r) &=& m_{\rm cs}R \,y_2 (m_{\rm cs}R) j_2 (m_{\rm cs}r),
\label{eqn:DsInt}
\end{eqnarray}
 and outside the sphere $(r\geq R)$
\begin{eqnarray}
    D_1(r) &=& \frac{2R}{r}\, j_2 (m_{\rm cs}R) y_1 (m_{\rm cs}r),  \nonumber\\          
    D_2(r) &=& m_{\rm cs}R\, j_2 (m_{\rm cs}R) y_1 (m_{\rm cs}r), \nonumber\\
    D_3(r) &=& m_{\rm cs}R \,j_2 (m_{\rm cs}R) y_2 (m_{\rm cs}r),
\label{eqn:DsExt}
\end{eqnarray}
where $j_\ell(x)$ and $y_\ell(x)$ are spherical Bessel functions of the first and second kind.  We see that the Chern-Simons terms alter the components of the gravitomagnetic field along the rotation axis $\vec{\omega}$ and $\hat r \times (\hat r \times \vec\omega)$, and they also introduce a new component perpendicular to the plane defined by $\vec{\omega}$ and $\vec{r}$.  In other words, while in general relativity a toroidal mass current implies a poloidal gravitomagnetic field, the parity violation introduced in Chern-Simons gravity introduces a toroidal component to the gravitomagnetic field.  Something similar occurs in Chern-Simons electromagnetism \cite{carroll}, although the detailed fields differ since the $\nabla^2 \vec B$ term in Eq.~(\ref{amp}) is simply $\vec B$ in the electromagnetic theory.

The Chern-Simons addition to Amp\`ere's law, Eq.~(\ref{amp}), changes that equation from a first-order differential equation for $\vec B$ to a second-order differential equation.  As a result, the Chern-Simons modification to the gravitomagnetic field cannot, in general, be obtained by perturbing around the general-relativistic result, as the solution in Eq.~(\ref{BfullCS}) shows.  In Chern-Simons gravity, the gravitomagnetic field oscillates with distance outside the source, and the amplitude of the oscillating field is not necessarily smaller than the general-relativistic gravitomagnetic field.  Still, we expect from Eq.~(\ref{amp}) that as $m_{\rm cs}\rightarrow\infty$, the general-relativistic solution should be recovered.  This occurs since the oscillatory terms vanish as $m_{\rm cs}\rightarrow\infty$, and so the effects on geodesics of these new terms will vanish.

If we take $\vec\omega$ to lie in the ${\hat z}$ direction, then the Chern-Simons gravitomagnetic field has a nonzero
azimuthal component $B_{\phi}$.  Since $B_{\phi}\neq0$, one cannot find a coordinate transformation that causes both $A_{r}$ and $A_{\theta}$ to vanish.  This is at odds with claims (see, e.g., Ref.~\cite{konno}) that
a metric for stationary axisymmetric spacetimes in Chern-Simons gravity can always be found with $h_{t\theta}=h_{tr}=0$.  In general relativity, one can always find a coordinate system for which $A_r = A_\theta=0$ for a stationary axisymmetric spacetime sourced by rotating perfect fluid.  However, the proof of this statement assumes time-reversal invariance of the fundamental equations.  This invariance implies that the metric components possess the same symmetries as the source, namely invariance under a transformation that takes $t \rightarrow -t$ and $\phi \rightarrow -\phi$.  In that case, $A_r$ and $A_{\theta}$ must be zero to keep the line element invariant under the same transformation.  In Chern-Simons gravity, time-reversal invariance is explicitly broken by the rolling of the scalar field, $\dot\theta\neq0$, and it is straightforward to verify that our solution for $\vec{A}$, given in Appendix B, implies that $A_r$ and $A_\theta$ are both odd under time reversal.  Consequently, the line element has the same symmetry as the source even though $A_r$ and $A_\theta$ are nonzero.

Inspection of our solution for the vector potential given in Appendix B shows that it differs from the solution for a point-like mass-current dipole (i.e., a gravitomagnetic dipole) obtained by Alexander and Yunes (AY) \cite{alexander}.  When applied to a single spinning source, the metric given by Refs.~\cite{alexander} corresponds to a vector potential
\begin{equation}
\vec{A}_{\rm AY} = \vec{A}_{\rm GR} -\frac{4 \pi G \rho R^3}{m_{\rm cs}R} \left[  \frac{2R^3}{15r^3} \, \vec\omega + \frac{R^3}{5r^3}\, \hat r \times (\hat r \times \vec\omega) \right] . \label{Aalex}
\end{equation}
This vector potential is an exact solution to Eq.~(\ref{amp}) outside of a spinning sphere, and we can see that every term in $\vec{A}_{\rm AY}$ also appears in our solution for $\vec{A}$.  The additional oscillatory terms in our solution constitute a homogeneous solution to Eq.~(\ref{amp}), but without these terms, $\vec{A}$ would not be continuous across the surface of the sphere.  Furthermore, only these oscillating terms contribute to $\vec{B}_{\rm CS}$ because $\vec\nabla \times \vec A_{\rm AY} = \vec\nabla \times \vec A_{\rm GR}$.  The inclusion of oscillatory terms results in a Chern-Simons gravitomagnetic field that differs from general relativity, so we may use observations of the motion of test bodies in the Earth's gravitomagnetic field to constrain Chern-Simons gravity.

\section{Orbital and Gyroscopic precession}

\subsection{Orbital precession}

In order to investigate how the Chern-Simons gravitomagnetic field will affect the motion of test particles around the Earth, we will
use what are known as the Gaussian perturbation equations \cite{brouwer,iorio}.  Details of how these equations are applied to gravitomagnetic forces are discussed in Ref.~\cite{smith}; here we give only a brief introduction.  The Gaussian perturbation equations give the time variation of the Keplerian orbital elements in the presence of a perturbing force.  In our case we take the gravitomagnetic force, $-4  \vec{v} \times \vec{B}$, as a small perturbing force and approximately solve the equations given in Ref.~\cite{smith}.  We will concentrate on analyzing the secular (non-periodic) time variation of the longitude of the ascending node\footnote{The longitude of the ascending node is defined to be the angle between a stationary reference line and the line connecting the origin of the coordinate system and the point where the orbiting body intersects the $XY$ reference plane as it is moving upwards (see Ref.~\cite{murray}).}, $\Omega$, but note that other Keplerian elements will also vary
due to the terms introduced by Chern-Simons gravity.
The time variation of $\Omega$ has been well studied since, in general relativity, it is connected with the Lense-Thirring drag \cite{lense_thirring},
\begin{equation}
     \dot{\Omega}_{\rm GR} = \frac{2 G L}{a^3(1-e^2)^{3/2}},
\end{equation}
where $L$ is the magnitude of the angular momentum of the central body, $a$ is the semi-major axis of the orbit of the test body, and $e$ is the orbit's eccentricity.  Finally, in order to evaluate the secular perturbations, we approximate the orbit of the test body as circular (i.e., $e=0$, a good approximation for current measurements), and we average the perturbing force over one orbital period to obtain
\begin{equation}
     \frac{\dot{\Omega}_{\rm CS}}{ \dot{\Omega}_{\rm GR} } = 15 \frac{a^2}{R^2}
     j_2(m_{\rm cs} R) y_1(m_{\rm cs} a),
\end{equation}
where $\dot{\Omega}_{\rm CS}$ is the precession due to $\vec{B}_{\rm CS}$.  The total precession is $\dot{\Omega}_{\rm GR}+\dot{\Omega}_{\rm CS}$.  We note that $\dot{\Omega}_{\rm CS}$ is an even function of $m_{\rm cs}$. 

\begin{figure}[h!]
\centerline{\epsfig{file=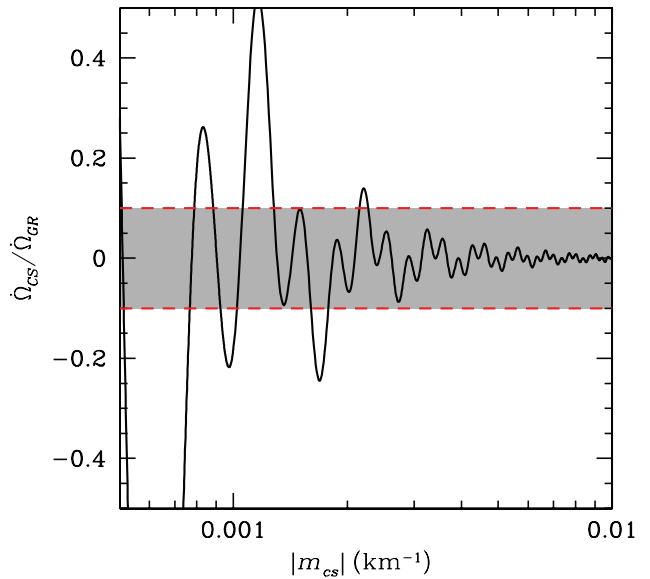,height=18pc,angle=0}}
\caption{The ratio
     $\dot{\Omega}_{\rm CS}/\dot{\Omega}_{\rm GR}$ for the LAGEOS
     satellites orbiting with a semimajor axis of $a \approx 12,
     000\ \mathrm{km}$.  A 10\% verification of general relativity
     \protect\cite{ciufolini} (the shaded region) leads to a lower
     limit on the Chern-Simons mass of $|m_{\rm cs}| \gtrsim 0.001\
     \mathrm{km}^{-1}$.  A 1\% verification of the
     Lense-Thirring drag will improve this bound on $m_{\rm cs}$ by
     a factor of roughly five.}
\label{fig:OmegaCSdot}
\end{figure}

Recent measurements of laser ranging data to the LAGEOS I and LAGEOS II satellites have measured  $\dot{\Omega}$ to within 10\% of its value in general relativity \cite{ciufolini}.  Requiring that the Chern-Simons contribution does not exceed 10\% of the general relativity result, we find that we can place a lower limit to the Chern-Simons mass, $|m_{\rm cs}| \gtrsim 0.001\ \mathrm{km}^{-1}$, as shown in Fig.~\ref{fig:OmegaCSdot}.  

The Laser Relativity Satellite (LARES) mission \cite{ciufolini2} proposes to deploy a new laser ranging satellite and is predicted to measure $\dot{\Omega}$ to within 1\% of its value in general relativity.  With this improvement the bound on $m_{\rm cs}$ is increased by a factor of roughly five.   

\subsection{Gyroscopic precession}

The Earth's gravitomagnetic field will also cause a precession of gyroscopes moving in the spacetime.  A gyroscope will undergo precession due to two torques.  One is known as the geodetic precession and is independent of the Earth's gravitomagnetic field.  The other torque is due to a coupling to the gravitomagnetic field and results in a rate of change of the spin of a gyroscope given by \cite{gyroscope}
\begin{equation}
     \dot{\vec{S}} = 2 \vec{B} \times \vec{S},
\label{LT_gyro}
\end{equation}
where $\vec{S}$ is the angular momentum of the gyroscope.

NASA's Gravity Probe B (GPB) mission is currently attempting to
measure this gyroscopic precession \cite{GPB}.   GPB consists of
a satellite, in a polar orbit at an altitude of about 640 km,
that contains four drag-free gyroscopes and a telescope.  The
gyroscopes are initially oriented such that their spins are
aligned parallel to the optical axis of the telescope, which is
pointing within the plane of the orbit.  The telescope points  
towards a guide star, allowing a measurement of the precession of
the direction of the spins of the gyroscopes.  Geodetic
precession results in an annual precession in the North-South
direction of about 6600 milliarcseconds (mas) whereas the general relativistic 
gravitomagnetic field causes an annual East-West precession of
around 42 mas \cite{GPB}.

\begin{figure}[!ht]
\centerline{\epsfig{file=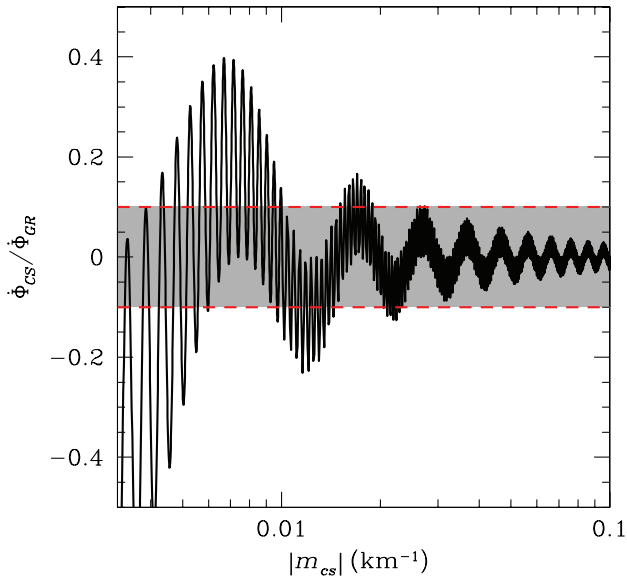,height=18pc,angle=0}}
\caption{The ratio
     $\dot{\Phi}_{\rm CS}/\dot{\Phi}_{\rm GR}$ for Gravity Probe B in
     a polar orbit at an altitude of approximately 640 km.
     A 10\% verification of general relativity (the shaded
     region) leads to a lower limit on the Chern-Simons mass of $|m_{\rm cs}|
     \gtrsim 0.01\ \mathrm{km}^{-1}$, an order of magnitude improvement over the LAGEOS
     result.}
\label{fig:GPB}
\end{figure}

With the Chern-Simons expression for the gravitomagnetic field, given in Eq.~(\ref{BfullCS}), it is straightforward to calculate the resulting
gyroscopic precession for a polar orbit (applicable to GPB).  Relative to the general relativity result, we find
\begin{eqnarray}
      \frac{\dot{\Phi}_{\rm CS}}{\dot{\Phi}_{\rm GR}} = 15
       \frac{a^2}{R^2}j_2(m_{\rm cs}R) \left[y_1(m_{\rm cs}a)+m_{\rm cs} a
        y_0(m_{\rm cs}a)\right],
\label{GPB_ratio}
\end{eqnarray}
where $\dot{\Phi} \equiv |\dot{\vec{S}}|/|\vec{S}| = \dot{\Phi}_{\rm GR} + \dot{\Phi}_{\rm CS}$ is the rate at which the angle of axis $\Phi$
changes in time due to the gravitomagnetic field.  We note that $\dot{\Phi}_{\rm CS}$ is an even function of $m_{\rm cs}$.

It was initially projected that GPB would achieve a
percent-level measurement of the gravitomagnetic contribution to $\dot{\Phi}_{\rm GR}$.  However, since
its launch in 2004, it has encountered several unexpected
complications that will degrade the precision of the tests of
gravity \cite{nature}, although the extent of the degradation
has yet to be reported.  In Fig.~\ref{fig:GPB}, we  plot
Eq.~(\ref{GPB_ratio}) for a GPB detection of the gravitomagnetic
precession to within 10\% of its value in general relativity.

We have idealized the Earth to be a sphere of constant density throughout this work, when in reality, it is an oblate spheriod with layers that have different mean densities.  However, we expect that the non-spherical corrections would affect both the general relativity and Chern-Simons calculations similarly and, to the accuracy we require, are negligible when we consider the ratio between general relativity and Chern-Simons results.  Furthermore, it is easy to generalize our results to spheres with layered density profiles because $\vec{B}$ depends linearly on $\rho$.  We replaced our model of a homogeneous Earth with a model of the core and mantle and we found that the amplitudes of the oscillations in $\dot{\Omega}_{\rm CS}$ and $\dot{\Phi}_{\rm CS}$ were not affected.  We conclude that our constraints on $m_{\rm cs}$ are not sensitive to the details of the density profile of the Earth.

\section{Conclusions}

The addition of a Chern-Simons term to the action for gravity
is of interest as it may arise as a low-energy limit of string
theory.  The theory and formalism of this modification of
gravity have been worked out in a number of previous papers, and
some of the early-Universe consequences of such a term have been
investigated.  However, there has been little work on tests of
such modifications in the present Universe.

In this paper, we have calculated the linear-theory
spacetime around a spinning massive body, finding new
corrections that were overlooked in previous work.  The
gravitomagnetic field in Chern-Simons gravity differs from that
in general relativity in two ways: (1) there is an oscillating
component, and (2) there is a toroidal component to the
gravitomagnetic field that arises as a consequence of the
parity-breaking nature of the theory and that has no counterpart
in ordinary general relativity.

We then determined the precession of orbits of test particles in
this spacetime and also of gyroscopes moving in this spacetime.
We showed that current constraints from the LAGEOS satellites
restrict the inverse Chern-Simons mass parameter $m_{\rm cs}^{-1}$
to be less than roughly $1000$ km, corresponding to a mass
constraint $m_{\rm cs} \gtrsim 2\times 10^{-22}$ GeV.  This bound may
be improved by a factor of 5-10 by future observations. 

The mass parameter $m_{\rm cs}$ is related to the more fundamental
parameters $\ell$ and $\dot \theta$ of the theory through
$m_{\rm cs} = -3/(8\pi G \ell \dot\theta)$, where $\ell$ is a length
parameter that enters into the Chern-Simons Lagrangian, and
$\dot\theta$ is presumably related to the time variation of the
quintessence field.  In principle, a precise constraint to
$\ell$ can be derived once the precise nature of the field (a
quintessence field?) $\theta$ and its time evolution are
specified.  We leave such model building for future work.

\acknowledgments

TLS thanks Ketan Vyas for useful conversations.  MK thanks the Aspen Center for Physics for hospitality during the completion of this work.  This work was supported at Caltech by DoE DE-FG03-92-ER40701,
NASA NNG05GF69G, and the Gordon and Betty Moore Foundation, and
by NSF AST-0349213 at Dartmouth.  ALE acknowledges the support of an NSF graduate fellowship. 

\appendix
\section{A string inspired derivation of the Chern-Simons field equations}
The effective 4-D string action for heterotic and type II string theory can be written as \cite{campbell,green}
\begin{equation}
     S = \int {\mathrm d}^4 x \sqrt{-g} \left[ -\frac{1}{2 \kappa^2} R -
     \alpha H_{\mu \nu \lambda}  H^{\mu \nu \lambda}
     + \cdots \right],
     \label{string_action}
\end{equation}
where $R$ is the Ricci scalar, $H_{\mu \nu \lambda}$ 
is the Kalb-Ramond (KR)
three-form field strength, and $\alpha$ is a constant with units of length squared.  We are neglecting numerous terms, including Gauss-Bonnet terms, 
dilaton terms, and matter terms, 
some of which depend on compactification.  The Kalb-Ramond field is written in
differential-form notation as
\begin{equation}
     \mbox{\boldmath $H$} = \frac{1}{3} \mathrm{d} \mbox{\boldmath $B$}  +
     \bomega_L, 
\end{equation}
where $\mbox{\boldmath $B$}$ is a two-form field (known as the
KR field) and $\bomega_L$ is the Lorentz-Chern-Simons term. 
 The Lorentz-Chern-Simons three-form can be
written in terms of the spin connection $\bomega$ as \cite{campbell3}
\begin{equation}
     (\bomega_L)_{\mu \nu \lambda} = \frac{1}{2}
     \mathrm{Tr}\left[ \omega_{[\lambda}(\mathrm{d} \omega)_{\mu \nu]} + \frac{4}{3} \omega_{[\mu}\omega_{[\nu}\omega_{\lambda]]}\right],
\end{equation}
where the trace is over the suppressed vector indices of the spin
connections.  We then have the identity,
\begin{equation}
     \mathrm{d}\mbox{\boldmath $H$}  =
     \frac{1}{6}\mathrm{Tr}(\mbox{\boldmath $R$}  \wedge  \mbox{\boldmath
     $R$} ),
\end{equation}
associated with the KR
field strength, where $\mbox{\boldmath $R$}$ is the Riemann
tensor and the
trace is over the tensor indices; the right-hand side is also
known as the Hirzebruch density.  Taking the Hodge dual of the
Hirzebruch density, we obtain 
\begin{equation}
     \frac{1}{6}\ ^* \mathrm{Tr} (\mbox{\boldmath $R$}  \wedge
     \mbox{\boldmath $R$} ) =  \frac{1}{4!} \epsilon^{\mu \nu
     \rho \lambda} R_{\alpha \beta \mu \nu} R^{\alpha  \beta}_{\
     \ \ \rho \lambda} = -\frac{1}{12}\RRdual.
\end{equation}

Let us now consider the equation of motion for the two-form KR
field.  We can rewrite the action involving $\mbox{\boldmath
$B$} $ as
\begin{eqnarray}
     S_B &\propto& \int \mbox{\boldmath $H$}  \wedge ^* \mbox{\boldmath $H$}- \bomega_L \wedge ^* \bomega_L \\
      &\propto& \int  \frac{1}{9}\mathrm{d} \mbox{\boldmath $B$}  \wedge ^*
      \mathrm{d}\mbox{\boldmath 
      $B$}  +\frac{1}{3} \mathrm{d}\mbox{\boldmath $B$}  \wedge ^*
      \bomega_L + \frac{1}{3}\bomega_L \wedge ^* \mathrm{d}\mbox{\boldmath
      $B$}  . \nonumber
\end{eqnarray}
On variation of this action with respect to $\mbox{\boldmath
$B$}$, we have the equation of motion,
\begin{equation}
     \mathrm{d}^* \mbox{\boldmath $H$}  = 0. 
\end{equation}
Therefore the equation of motion for the KR two-form field shows that $^* 
\mbox{\boldmath $H$} $ is closed.  In other words, at least locally,
 there exists a pseudo-scalar $b$ (the KR axion, or
sometimes called the universal axion) such that
\begin{equation}
     \mbox{\boldmath $H$}  =\ ^* \mathrm{d}b.
\end{equation}
Noting that $-^*\mathrm{d}^*\mathrm{d} \phi = \Box \phi$, we
have the equation of motion for $b$,
\begin{equation}
     \Box b=-  ^* \mathrm{d} \mbox{\boldmath $H$}  =-\frac{1}{6}\ ^*
     \mathrm{Tr} (\mbox{\boldmath 
     $R$}  \wedge \mbox{\boldmath $R$} ) = \frac{1}{12} \RRdual.
\end{equation}

Varying the action given by Eq.~(\ref{string_action}) with respect to the metric
we obtain\footnote{The sign of the last term in this equation is different in Ref.~\cite{campbell2}.  The sign given here makes the divergence of the right-hand side vanish as required by the Bianchi identity.} \cite{campbell2}
\begin{eqnarray}
     -G^{\mu \nu} = &&\kappa^2 \alpha \{6 H^{\mu}_{\ \  \lambda
     \rho} H^{\nu \lambda  \rho} - g^{\mu \nu} H^{\lambda \rho
     \sigma} H_{\lambda \rho \sigma} \\ 
     &&+ 4\nabla_{\sigma}(H^{\lambda \alpha (\mu} R^{\nu)
     \sigma}_{\ \ \ \ \alpha  \lambda})\}, \nonumber
 \end{eqnarray}
where $G^{\mu \nu}$ is the usual Einstein tensor. 
Given that the equation of motion for the two-form field
$\mbox{\boldmath $B$} $ allows us to write $\mbox{\boldmath $H$}
=\ ^* \mathrm{d} b$, we have 
\begin{equation}
     H_{\mu \nu \rho} = \epsilon^{\sigma}_{\ \mu \nu \rho}
     \nabla_{\sigma} b.
\end{equation}
We can rewrite the field equation as
\begin{equation}
     -G^{\mu \nu} = \kappa^2 \alpha 12 \left[T_b^{\mu \nu} +
     \frac{1}{3}\nabla_{\sigma}(H^{\lambda \alpha (\mu} R^{\nu)
     \sigma}_{\ \ \ \ \alpha  \lambda})\right],
\end{equation}
where $T_b^{\mu \nu}$ is the canonical stress-energy tensor for
the pseudo-scalar field $b$. 
We will now show that the last term is actually the Cotton-York
tensor.

Using the Bianchi identities for the Riemann tensor, we first
note that we have the identity,
\begin{equation}
     \nabla_{\sigma} \tilde R^{\sigma (\mu| \tau |\nu)} =
     \epsilon^{(\mu|\tau \sigma \rho}  \nabla_{\rho}
     R^{|\nu)}_{\ \ \sigma}.
\label{identity}
\end{equation}
With this, it is straightforward to show that
\begin{eqnarray}
     && \nabla_{\sigma}([\nabla_{\tau} b] \epsilon^{\tau \lambda
     \alpha (\mu} R^{\nu)  \sigma}_{\ \ \ \ \alpha \lambda})
     \nonumber \\ 
      &&= 2 \nabla_{\sigma}([\nabla_{\tau} b]
     \tilde{R}^{\sigma(\nu| \tau| \mu)}) = 2  C^{\mu \nu},
\end{eqnarray}
where $C^{\mu \nu}$ is the Cotton-York tensor defined in Eq.~(\ref{cy}). 
Choosing $\alpha = \ell^2/12$ and taking $b \rightarrow -\theta/\ell$
so that in the absence of the Cotton-York tensor we regain general relativity
sourced by a canonical scalar field $\theta$, the equations of motion are 
\begin{eqnarray}
     G^{\mu \nu}-\frac{2 \ell \kappa^2}{3} C^{\mu \nu} &=&
     -\kappa^2 T_{\theta}^{\mu \nu}, \\ 
     \Box \theta &=&  -\frac{1}{12} \ell \RRdual.
\end{eqnarray}
We can see that these field equations are identical to
Eqs.~(\ref{scalar}) and (\ref{field_equation}) with vanishing
scalar potential.  

\section{Calculation of the vector potential}
In Lorenz gauge ($\partial_{\mu} A^{\mu} = 0$) the Chern-Simons
Amp\`ere's law, Eq.~(\ref{amp}), can be written,
\begin{equation}
     \Box \left[\vec{A}+ \frac{1}{m_{\rm cs}}\vec{B}\right] = -4 \pi
     G \vec{J},
\label{eqn:altamp}
\end{equation}
where we have neglected the time variation in $\dot{\theta}$ in order to 
place $m_{\rm cs}$ inside the d'Alembertian operator. 
We are dealing with a stationary source, and so $\Box = 
\nabla^2$.  We may invert Eq.~(\ref{eqn:altamp}) to obtain
\begin{equation}
     \vec{A} + \frac{1}{m_{\rm cs}} \vec{\nabla} \times \vec{A} = G \int 
     \frac{\vec{J}}{|\vec{r} - \vec{r}^{\prime}|} {\mathrm d}^3 r^{\prime}.
\label{eq:cool!}
\end{equation}
We can write this as
\begin{equation}
     \left(\mathcal{I} +\frac{1}{m_{\rm cs}} \vec{\nabla} \times
     \right)\vec{A} = G \int \frac{\vec{J}}{|\vec{r} -
     \vec{r}^{\prime}|} {\mathrm d}^3 r^{\prime},
\end{equation}
where $\mathcal{I}$ is the identity matrix.  Multiplying both
sides of the equation by $\left[\mathcal{I} -(1/m_{\rm cs}) \vec{\nabla}
\times\right]$, we obtain
\begin{equation}
     \vec{A} -\frac{1}{m_{\rm cs}^2} \vec{\nabla} \times \vec{\nabla}
     \times \vec{A} =  G \left(\mathcal{I} -\frac{1}{m_{\rm cs}}
     \vec{\nabla} \times \right) \int \frac{\vec{J}}{|\vec{r}  -
     \vec{r}^{\prime}|} {\mathrm d}^3 r^{\prime}.
\end{equation}
Noting that $\vec{\nabla} \times \vec{\nabla} \times \vec{A} = -
\nabla^2 \vec{A}$ in Lorenz gauge, we have
\begin{equation}
      \nabla^2 \vec{A} + m_{\rm cs}^2\vec{A}= \vec S,
\label{diffeq}
\end{equation}
where 
\begin{equation}
     \vec S \equiv m_{\rm cs}^2 G \left(\mathcal{I} -\frac{1}{m_{\rm cs}}
     \vec{\nabla} \times \right)  \int \frac{\vec{J}}{|\vec{r} -
     \vec{r}^{\prime}|} {\mathrm d}^3 r^{\prime}.
\label{eq:source}
\end{equation}
We recognize this as the inhomogeneous Helmholtz equation. 
For a rotating homogeneous sphere, the mass current is given by
\begin{equation}
     \vec{J} = \rho [\vec{\omega} \times \vec{r}] \Theta(R-r),
\end{equation}
where $\rho$ is the density, $\omega$ is the angular velocity,
$R$ is the radius, and $\Theta$ is the Heaviside step function. 

The most general Green's function for the inhomogeneous Helmholtz
equation is
\begin{equation}
     G(\vec{r},\vec{r}^{\prime}) =
     -\frac{\cos(m_{\rm cs}|\vec{r}-\vec{r}'|) + \tilde{\gamma}
     \sin(m_{\rm cs}|\vec{r}-\vec{r}'|)}{4\pi|\vec{r}-\vec{r}'|},
\end{equation}
where $\tilde \gamma$ is a constant.  However, the second term
(that is proportional to $\tilde \gamma$) remains constant for
$|\vec r'-\vec r|\ll m_{\rm cs}^{-1}$, implying that the influence of the
source does not decrease with distance (for distances $r \ll
m_{\rm cs}^{-1}$), which we interpret as unphysical.  We therefore
set $\tilde\gamma=0$.  We then use multipole expansions for the Green's function,
\begin{eqnarray}
    && -\frac{\cos(m_{\rm cs}|\vec r - \vec r'|)}{ 4 \pi |\vec r - \vec r'|} = \\
    && m_{\rm cs}\sum_{\ell, m} j_\ell(m_{\rm cs} r_<) y_\ell(m_{\rm cs}
     r_>)Y^*_{\ell m}(\hat r') Y_{\ell m}(\hat r),\nonumber
\label{eqn:multexpansions}
\end{eqnarray}
where $j_{\ell}(x)$ and $y_\ell(x)$ are, respectively, spherical
Bessel function of the first and second kind,
$Y_{\ell m}(\hat{r})$ is a spherical harmonic, and the
subscript $<$ ($>$) means the argument is the lesser (greater)
of $r$ or $r'$.  The solution for $\vec{A}$ is
then obtained by integrating,
\begin{equation}
     \vec{A} = \int {\mathrm d}^3 r^\prime G(\vec{r},{\vec{r}}^{\prime}) 
     \vec{S}({\vec{r}}^{\prime}),
\end{equation}
where all vectors are expanded in a Cartesian basis. 

The resulting expression for $\vec{A}$ may be split into a
general-relativistic and a Chern-Simons term, $\vec{A} 
= \vec{A}_{\rm GR} + \vec{A}_{\rm CS}$, where
\begin{equation}
     \vec{A}_{\rm GR} = -\frac{4\pi G \rho}{3} R^3 ( \hat{r} \times
     \vec{\omega})\times
\begin{cases}
        \frac{r}{R}\left[\frac{1}{2} -
        \frac{3}{10}\left(\frac{r}{R}\right)^2\right], \ &r
        \leq R, \\
        \frac{R^2}{5r^2}, \ &r \geq R,
\end{cases}
\label{AfullGR}
\end{equation}
is the gravitomagnetic vector potential in general relativity, and 
\begin{eqnarray}
     \vec A_{\rm CS} = -\frac{4 \pi G \rho R^3}{m_{\rm cs}R} &&\left[
     C_1(r)\, \vec\omega +   C_2(r)\, \hat r \times \vec\omega
     \right. \nonumber \\
     &&\left. + C_3(r) \, \hat r \times (\hat r \times
     \vec\omega) \right],
\label{AfullCS}
\end{eqnarray}
with
\begin{eqnarray}
     C_1(r) &=&  -\frac{r^2}{5R^2} +\frac{1}{3} + 
     \frac{2}{m_{\rm cs}^2R^2}+\frac{2R}{r} \, y_2 (m_{\rm cs}R) j_1 (m_{\rm cs}r),
     \nonumber\\
     C_2(r) &=& \frac{m_{\rm cs} r}{m_{\rm cs}^2 R^2}+m_{\rm cs}R \, y_2 
     (m_{\rm cs}R) j_1 (m_{\rm cs}r),\nonumber \\
     C_3(r) &=& \frac{r^2}{5R^2}+m_{\rm cs}R \, y_2 (m_{\rm cs}R) j_2 
     (m_{\rm cs}r),
\end{eqnarray}
inside the sphere, and
\begin{eqnarray}
     C_1(r) &=& \frac{2R^3}{15r^3}+\frac{2R}{r}\, j_2 (m_{\rm cs}R)
     y_1 (m_{\rm cs}r), \nonumber\\
     C_2(r) &=& m_{\rm cs}R\, j_2 (m_{\rm cs}R) y_1 (m_{\rm cs}r), \nonumber\\
     C_3(r) &=& \frac{R^3}{5r^3}+m_{\rm cs}R \,j_2 (m_{\rm cs}R) y_2
     (m_{\rm cs}r), \nonumber\\
\end{eqnarray}
outside the sphere.  We note that this solution for $\vec{A}$ is
finite at the origin and continuous across the boundary of the
sphere, so it produces a finite $\vec{B}$ at the origin and a
continuous metric.  Taking the curl of this solution for $\vec{A}$ yields the expressions for $\vec{B}$ given in Section \ref{sec:Bfield}.

Thus far, we have not discussed any boundary conditions on the gravitomagnetic field $\vec{B}$ at the surface of the sphere.  The field equations for $\vec{B}$ imply two such boundary conditions, and we will now prove that the continuity of $\vec{A}$ guarantees that these two boundary conditions are satisfied.  The first boundary condition follows from $\vec\nabla \cdot \vec{B} = 0$; as in electromagnetism, this condition implies that the component of $\vec{B}$ that is perpendicular to the surface must be continuous.  The second boundary condition follows from the Chern-Simons version of Amp\`ere's law:
\begin{equation}
\vec\nabla \times \vec{B} - \frac{1}{m_{\rm cs}} \nabla^2 \vec{B}  = 4 \pi G \vec{J}.
\end{equation}
Integrating this equation over a surface with vanishing area that is perpendicular to the surface of the sphere and contains the boundary implies that the components of $[\vec{B} + (1/m_{\rm cs})\vec{\nabla}\times \vec B]$ that are parallel to the sphere's surface must be continuous across the boundary.

Generally, the continuity of $\vec{A}$ would not imply continuity of its curl.  However, our $\vec{A}$ is a solution to  Eq.~(\ref{eq:cool!}), which may be rewritten as
\begin{equation}
\vec{A} + \frac{1}{m_{\rm cs}} \vec{B}  = \vec{A}_{\rm GR}.
\end{equation}
Since $\vec{A}$ and $\vec{A}_{\rm GR}$ are both continuous across the surface of the sphere, this equation implies that $\vec{B}$ is also continuous across the surface of the sphere.  Furthermore, taking the curl of this equation shows that $\vec \nabla \times \vec B$ is continuous provided that $\vec B$ and  $\vec \nabla \times \vec{A}_{\rm GR}$ are continuous.  Taking the curl of Eq.~(\ref{AfullGR}) confirms that $\vec \nabla \times \vec{A}_{\rm GR}$ is continuous across the surface of the sphere.  Therefore, we have shown that the continuity of $\vec{A}$ implies that both $\vec{B}$ and $\vec \nabla \times \vec B$ are also continuous, which guarantees that both boundary conditions on $\vec{B}$ are satisfied by our solution.


\end{document}